\documentclass[fleqn,twoside,twocolumn,nofootinbib]{revtex4} 
\usepackage[sec]{ujp} 
\begin{document}
\title[CONFLICTING COUPLING OF UNPAIRED NUCLEONS]
{CONFLICTING COUPLING OF UNPAIRED NUCLEONS AND
THE STRUCTURE OF COLLECTIVE BANDS\\ IN ODD-ODD NUCLEI}%
\author{A.I. LEVON}
\affiliation{Institute for Nuclear Research, Nat. Acad. of Sci. of Ukraine}
\address{47, Prosp. Nauky, Kyiv 03680, Ukraine}
\author{A.A. PASTERNAK}%
\affiliation{A.F. Joffe Physical-Technical Institute, Acad. of Sci.
of Russia
}%
\address{St. Petersburg, Russia}%
\udk{533.9} \pacs{21.10.-k, 21.60.-n,\\[-3pt] 25.40.Hs, 27.90.+b}
\razd{\secii}

\setcounter{page}{107}%
\maketitle

\begin{abstract}
The conflicting coupling of unpaired nucleons in odd-odd nuclei is
discussed. A very simple explanation is suggested for the damping of
the energy spacing of the lowest levels in the rotational bands in
odd-odd nuclei with the ``conflicting'' coupling of an odd proton
and an odd neutron comparative to those of the bands based on the
state of a strongly coupled particle in the neighboring odd nucleus
entering the ``conflicting'' configuration.
\end{abstract}

 \section{Introduction}
There are two coupling schemes of the unpaired nucleons in odd-odd
nuclei referring to two different mutual orientations of their
angular momenta. If the unpaired neutron and proton are coupled to
the deformed core in the same way (strongly coupled or decoupled),
i.e. the angular momenta ${\bf j}_{n}$  and $ {\bf j}_{p}$ are both
oriented either along the symmetry axis or along the rotation axis,
the situation is termed ``peaceful'' coupling [1] of the unpaired
nucleons to the deformed core. In this case, the structure of the
collective bands can adequately be described in terms of the model
``axial rotor + quasiparticle'' if the proton and the neutron are
considered as a single ``superquasiparticle'' [2]. If the neutron
and the proton are coupled to the core, with the angular momentum of one
nucleon being oriented along the symmetry axis and the other along
the rotation axis (Fig. 1), the  coupling of the unpaired nucleons
with the deformed even-even core is referred to as ``conflicting''
[3]. In the case of a prolate (oblate) deformation, the ``conflicting''
coupling can be realized in nuclei, in which the nucleons of one kind
just start to fill the Nilsson orbits belonging to a definite
single-particle  $lj,$
 decoupled (strong coupled), and the
orbits of the nucleons of the other kind are almost filled, strong
coupled (decoupled). Such a situation can be found in the regions of
nuclear masses given in Table~\ref{tab:regions}.\looseness=1

\begin{table}[b]
\noindent\caption{\label{tab:regions}Regions of nuclear masses,
where the ``conflicting'' coupling of unpaired nucleons in
odd-odd nuclei can be realized }\vskip3mm\tabcolsep9.7pt

\noindent{\footnotesize\begin{tabular}{c c c c }
 \hline \multicolumn{1}{c}
{\rule{0pt}{9pt}Mass} & \multicolumn{1}{|c}{Shape}&
\multicolumn{2}{|c}{Orbits of
  definite single-particle $lj$}\\ \cline{3-4}%
\multicolumn{1}{c}{of nuclei}& \multicolumn{1}{|c}{}&
\multicolumn{1}{|c}{beginning to fill}& \multicolumn{1}{|c}{almost
closed }\\
\hline%
$\sim 110$& prolate &$\nu h_{11/2}$&$\pi g_{9/2}$\\
$\sim 135$& prolate &$\pi h_{11/2}$&$\nu h_{11/2}$\\
$\sim 170$& prolate &$\nu i_{13/2}$&$\pi h_{11/2}$\\
$\sim 200$& oblate  &$\pi h_{9/2}$ &$\nu i_{13/2}$\\
\hline
\end{tabular}}
\end{table}

\begin{figure}
\includegraphics[width=7.5cm]{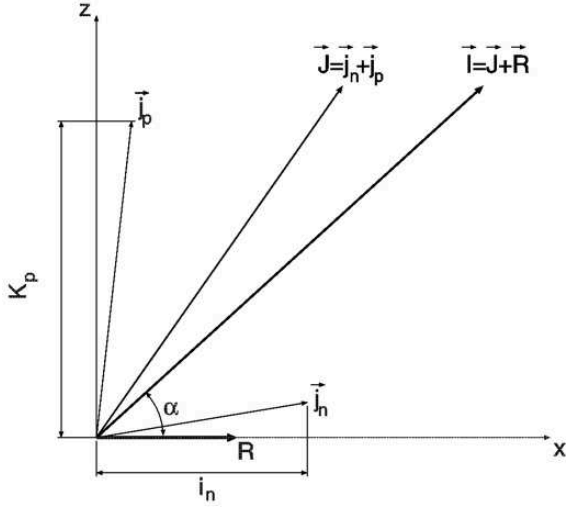}
\caption{ Possible  type of coupling of the odd nucleons to the
deformed core in odd-odd nuclei: the ``conflicting'' coupling }
\end{figure}

One of the manifestations of the ``conflicting'' coupling was
considered in our paper \cite{Lev92}: the reduced M1-transition
probabilities between collective-band levels in the odd-odd nuclei
are enhanced in comparison with those in the neighboring odd nuclei
with strong coupling of the odd nucleon. The aim of the present
paper is to investigate the structure of the collective bands in
those odd-odd nuclei that are based on the ``conflicting'' states of
the unpaired proton and neutron. Particularly, we consider the reduction of the
energy intervals at the beginning of the ``conflicting'' bands
relative to the bands in the neighboring odd nuclei using a
quasiclassical limit of the model ``axial rotor + two
quasiparticles'' in the analysis.

\section{Theoretical Background}

The Hamiltonian of an odd-odd nucleus in the model ``axial rotor +
two quasiparticles'' can be written as
\begin{equation}
 \hat{H} = \hat{H}_{p} + \hat{H}_{n} + A \hat{R}^{2} + \hat{V}_{p-n},
\end{equation}
where $A=\hbar^{2}/2{\cal J}$ is the inertial parameter,
$\hat{H}_{p}$ and $\hat{H}_{n}$ are the single-particle Hamiltonians
for a proton and a neutron, $\hat{V}_{n-p}$ is the residual $n$-$p$ --
interaction, and $\hat{R}$ is the operator of collective
 rotation. Since ${\bf R} =
{\bf I} - ( {\bf j}_{p} + {\bf j}_{n} ),$ where ${\bf I}$ is the
total angular momentum of the nucleus, and ${\bf j}_{p}$ and
  ${\bf j}_{n}$ are the angular momenta of the odd proton
and neutron, respectively, expression (1) can be written as
\[\hat{H} = \hat{H}_{p} + \hat{H}_{n} + \]
\begin{equation}\label{1}
+ A \left\{ \hat{I}^{2} - \hat{I}_{z}^{2} - 2
(\hat{I}_{+}\hat{J}_{-} +\hat{I}_{-}\hat{J}_{+}) + (\hat{J}^{2} -
\hat{J}_{z}^{2})\right\} + \hat{V}_{n-p},
\end{equation}
where ${\bf J} = {\bf j}_{p} + {\bf j}_{n}.$ The term
$2A(\hat{I}_{+}\hat{J}_{-} + \hat{I}_{-}\hat{J}_{+})$ represents the
Coriolis interaction of the odd nucleons with the rotating core, and
$\hat{I}_{\pm} =\hat{I}_{x} \pm i\hat{I}_{y}$ and $\hat{J}_{\pm} =
\hat{J}_{x} \pm i\hat{J}_{y}.$ The term $A(\hat{J}^{2} -
\hat{J}_{z}^{2})$  represents the recoil energy of the rotor which
depends only on the quantum numbers of the odd nucleons.

The contribution of the term $\hat{V}_{n-p}$ is the
most difficult one to estimate. However, as the matrix elements of the operator
$\hat{V}_{n-p}$ do not depend directly on $I,$ they are much smaller
 for the high-spin states than the matrix elements of the Coriolis interaction.
Furthermore, it is possible to write  \cite{Jai89}
\begin{equation}
\langle IMK \Omega_{p}\Omega_{n} \mid \hat{V}_{n-p} \mid IMK
\Omega_{p}\Omega_{n}\rangle = E_{o} + (-)^{I}B,
\end{equation}
where $K,$ $\Omega_{p},$ and $\Omega_{n}$ are the projections on the
symmetry axis of $J,$ $j_{p},$ and $j_{n},$ respectively. The second
term in (3) leads to the odd-spin even-spin shift of the levels of
the rotational band, but it is different from zero only if $K=0,$
that is when $\Omega_{p} = -\Omega_{n} .$ Since, in the case of
``conflicting'' coupling, the Fermi level of a decoupled nucleon is
close to the orbitals with the lowest value of the angular momentum
projection on the symmetry axis $\Omega=1/2,$ and the Fermi level of
the strongly coupled nucleon is close to the orbitals with the
highest value of the projection $\Omega=j,$ the condition
$\Omega_{p} = -\Omega_{n} $ is never fulfilled. Therefore, only the
first term in (3) is present, by leading to equal shifts of all levels
of the collective band and thus
 not influencing its structure. As we are interested only in the structure
 of the collective bands, i.e. in the energies of the band levels relative to
the energy of the band head, this constant shift of all levels is ignored.
It has no influence on the band structure, but it is important for the
absolute values of the level energies.

Taking these arguments about $\hat{V}_{n-p}$ into account, the
specific properties
of the rotational bands in odd-odd nuclei are determined by the
diagonalization of the Coriolis interaction of the odd nucleons
with the rotating core [4]. Therefore, the rotational part of (2) can be
written as
\begin{equation}
\hat{H}_{\rm rot}=A[\hat{I}^{2}-\hat{I}_{z}^{2}-
2(\hat{I}_{+}\hat{J}_{-}+\hat{I}_{-}\hat{J}_{+})+
  (\hat{J}^{2}-\hat{J}_{z}^{2})].
\end{equation}

We now consider the case of extreme ``conflicting'' coupling, i.e.
one nucleon is completely decoupled, and the other is completely
coupled to the deformed even-even core. This corresponds to the
classical limit where the angle between the vectors
 ${\bf j}_{p}$ and  ${\bf j}_{n}$ is
equal to $90^{\circ},$ and a weak ``conflicting'' coupling
corresponds to
 angles less than $90^{\circ}.$ Let us suppose  that the odd
neutron is a decoupled particle and the odd proton is a
strongly coupled particle. Then the proton state is characterized by
the maximum value
of $\Omega_{p} = j_{p}.$ The neutron is oriented along
the rotation axis, and the neutron state is characterized by the
minimum value of $\Omega_{n}.$

As a measure for the decoupling or alignment, Flaum and Cline
\cite{Fla76} proposed to take the expectation value of the
projection of the intrinsic angular momentum ${\bf J}$ onto the
component of the total angular momentum  ${\bf I}$ perpendicular to
the symmetry axis,
\begin{equation}
J_{\perp} =\left\langle \frac{{\bf I} \cdot {\bf J} -
K^{2}}{I_{x}}\right\rangle =
\left\langle\frac{I_{+}J_{-}+I_{-}J_{+}}{[(I+1)I -
K^{2}]^{1/2}}\right\rangle,
\end{equation}
where $ I_{x} =\sqrt{ (I+1)I-K^{2}} $ is the projection of the total
angular momentum on the rotation axis. The choice of (5) as a measure
for the alignment is convenient, because $J_{\perp}=0$ for the strong
coupling of both odd nucleons. In this case, ${\bf I} \cdot {\bf J} =
K^{2} ,$ which corresponds to the orientation of
${\bf J}$ along the symmetry axis. In the case of
the full decoupling, $J_{\perp}=J,$ because $ I_{+}J_{-}+I_{-}J_{+} =
JI_{x} $  \cite{Ste75},
which corresponds to the orientation of
${\bf J}$ along the rotation axis.

In view of (5), the rotational part of Hamiltonian (1)
can be rewritten as
\begin{equation}
\langle\hat{H}_{\rm rot}\rangle =
A[(I_{x}-J_{\perp})^{2}-J_{\perp}^{2}+
\langle\hat{J}^{2}-\hat{J}_{z}^{2}\rangle].
\end{equation}
It is possible to write
\begin{equation}
{\bf I}_{\rm odd-odd}={\bf R}+{\bf j}_{n}+{\bf j}_{p}={\bf I}_{\rm
odd}+{\bf j}_{n},
\end{equation}
where ${\bf I}_{\rm odd}={\bf R}+{\bf j}_{p}$ is the total angular
momentum of the levels of a rotational band in the neighboring
odd nucleus, based on the state of a strongly coupled proton.

Therefore, in the case under consideration, we have $ (I_{x}-J_{\perp})_{\rm
odd-odd}=(I_{x})_{\rm odd}.$ Since $ (I_{x}^{2})_{\rm odd}=I_{\rm
odd}(I_{\rm odd}+1)-K_{p}^{2},$ $K_{p}=j_{p}$ for a strongly coupled proton,
and $ J_{\perp}=j_{n}$ for the extreme ``conflicting'' coupling with
the neutron being decoupled,
expression (6) takes the form
\begin{equation}
\langle\hat{H}_{\rm rot}\rangle = A[I_{\rm odd}(I_{\rm
odd}+1)-K^{2}],
 \end{equation}
where $ K=\Omega_{p}+1/2 $ for the  case considered in our example.

Thus, the energy spectrum of the rotational band in an odd-odd
nucleus based on the ``conflicting'' state is determined by the
rotational excitations of the neighboring odd nucleus with the
strongly coupled particle being the one entering the ``conflicting''
state in the odd-odd nucleus, where a sequence of levels with
$\Delta I=1$ is observed. The spin value of such a band head is
approximately equal to $ I_{\rm head}=\sqrt{j_{n}^{2}+j_{p}^{2}}.$
In the quasiclassical limit, $ \parallel I\parallel =
\parallel {\bf j}_{n}+{\bf j}_{p} \parallel =
\sqrt{j_{n}^{2}+j_{p}^{2}+ 2j_{p}j_{n}\cos({\bf j}_{n},{\bf j}_{p})}
$ with $ \cos({\bf j}_{n},{\bf j}_{p})=0 $ in the case of
``conflicting'' coupling.

There is a correspondence between the ``conflicting'' bands in
the odd-odd nuclei and the rotation-aligned bands
in the odd nuclei. In the rotation-aligned bands, the angular momentum of the
odd nucleon does not influence the structure of the
rotational band which is similar to that of the neighboring
even-even nucleus, and the decoupled nucleon can be considered as a
spectator. In the case of  ``conflicting'' bands in  odd-odd nuclei,
the decoupled particle can be considered as a spectator, but its
role is somewhat more complicated, as will be explained in
the following chapter.

\section{The Structure of the Collective Bands in Odd-Odd Nuclei
in the Case of  Extreme ``Conflicting'' Coupling}

The analysis of the energy spectra of  rotational bands in odd-odd
nuclei investigated experimentally by the in-beam $\gamma$-spectroscopy
allowed one to identify ``conflicting'' bands in the following nuclei:
$^{102,104,106}$Ag \cite{Pop81}, $^{106,108,110}$In
\cite{Wis81,Eli81,Ber89}, $^{112,114,116,118,120}$Sb
\cite{Qud86,Van82,Vaj83}, $^{116,118,120,122}$I \cite{Qua84}, and
$^{120,122}$Cs \cite{Qua86} with a decoupled h$_{11/2}$-neutron and
a strongly coupled g$_{9/2}$-proton; $^{124,128}$Cs
\cite{Kom90,Pau89}, $^{126,128,130,132}$La
\cite{Pau87,God89,Nay89,Oli89}, $^{130,132,134}$Pr
\cite{Ma88,Shi88,Bea87}, $^{136}$Pm \cite{Bea87}, and $^{138}$Eu
\cite{Lia88} with a decoupled h$_{11/2}$-proton and a
strongly coupled h$_{11/2}$- or g$_{7/2}$-neutron; $^{170}$Ta
\cite{Bac85} and $^{194,196,198,200}$Tl \cite{Kre79,Kre77} with a
decoupled i$_{13/2}$-neutron and a strongly coupled
h$_{11/2}$-proton. Common to all these nuclei is a well-developed
band with a $\Delta I=1$ sequence built on states with spins
approximately equal to $\sqrt{j_{p}^{2}+j_{n}^{2}}.$

Let us compare the energy intervals between the levels of
rotational bands in the odd-odd  and in the neighboring odd
nuclei. According to (8), the spectrum in an odd-odd nucleus for
the case of ``conflicting'' coupling is determined by the total
angular momentum of the odd nucleus. Therefore, the difference in
the energy intervals has to be caused by a modification of the
effective moment of inertia going from the odd to the odd-odd
nucleus.

We assume that the dependence of the effective moment of inertia
on the rotation frequency for a band based on the
``conflicting'' state can be expressed as
\begin{equation}
{\cal J}_{\rm odd-odd}(\omega)={\cal J}_{\rm
odd}(\omega)+\alpha/\omega,
\end{equation}
where the first term is the moment of inertia of the rotating
deformed core, the second term is the contribution of the angular
momentum of the decoupled particle ($\alpha$ is the alignment of
the decoupled particle). This expression originated from the
cranking model \cite{Ben79}, but it can also be considered as a
reflection of the assumption about an additional contribution to the
moment of inertia based on the known relation  \( I_{x}={\cal
J}\omega \). It is simply based on the additivity of the angular
momentum projections of the core and the decoupled particle.
Applying this relation to the moment of inertia of an odd-odd
nucleus, expression (9) can be transformed to
\begin{equation}
{\cal J}_{\rm odd-odd}(\omega)={\cal J}_{\rm odd}(\omega)
  (1-\alpha/(I_{x})_{\rm odd-odd})^{-1}.
\end{equation}
The alignment $\alpha$ is defined as
 \( \alpha=(I_{x})_{\rm odd-odd}-(I_{x})_{\rm odd} \),
where $(I_{x})_{\rm odd}$ is the projection of the total angular
momentum of the odd nucleus  on the rotation axis. For the moment of
inertia, we find
\begin{equation}
{\cal J}_{\rm odd-odd}(\omega)={\cal J}_{\rm odd}(\omega)
 \frac{(I_{x})_{\rm odd}}{(I_{x})_{\rm odd-odd}}.
\end{equation}

Then we notice that the rotational band in an odd-odd nucleus can
be represented in the usual form
\begin{equation}
 E_{\rm odd-odd} = \frac {\hbar^{2}} {2{\cal J}_{\rm odd-odd}}
  [I_{\rm odd-odd}(I_{\rm odd-odd}+1)-K^{2}]
\end{equation}
and, according to (8), in the form
\begin{equation}
 E_{\rm odd-odd}^{\prime} = \frac {\hbar^{2}} {2{\cal J}_{\rm odd-odd}^{\prime}}
  [I_{\rm odd}(I_{\rm odd}+1)-K^{\prime 2}].
\end{equation}
The relation between the two moments of inertia ${\cal J}_{\rm
odd-odd}$ and ${\cal J}_{\rm odd-odd}^{\prime}$ is obtained by
comparing the energy intervals between the levels with spins $I$ and
$I-1$ calculated with Eqs. (12) and (13)
\begin{equation}
{\cal J}_{\rm odd-odd}(\omega)={\cal J}_{\rm
odd-odd}^{\prime}(\omega) \frac{I_{\rm odd-odd}}{I_{\rm odd}}.
\end{equation}
Then, using (11) and (14) with regard for the relation $I_{x}=I \sin
\theta,$ where $\theta$ is the angle between the vector of the total
angular momentum and the symmetry axis so that
\begin{equation}
\sin \theta = [1-(K/I)^{2}]^{1/2}
\end{equation}
in the quasiclassical approximation, we represent the moment of
inertia in the form
\begin{equation}
{\cal J}_{\rm odd-odd}^{\prime}(\omega)={\cal J}_{\rm odd}
 \frac{\sin \theta_{\rm odd}}{\sin \theta_{\rm odd-odd}}.
\end{equation}

The geometric factor $ \sin \theta_{\rm odd-odd}/ \sin \theta_{\rm
odd}$ which is determined by the orientation of the total angular
momenta in the odd-odd and odd nuclei leads to the relation \(
{\cal J}_{\rm odd-odd}^{\prime} > {\cal J}_{\rm odd} \) for the
low-lying states of the rotational bands. As is evident from (13),
this leads to a reduction of the lowest energy intervals in the
rotational band of the odd-odd nucleus compared to the corresponding
intervals of the reference band in the odd nucleus. The geometric
factor approaches 1 for the high-lying states, and the corresponding
energy intervals turn out to be almost equal.

The same result can be obtained by comparing the energy intervals
$\Delta E(I,I-1)$ in the odd-odd and odd nuclei:
\begin{equation}
{\Delta E}_{\rm odd-odd}= \frac{{\hbar}^{2}}{2{ \cal J}_{\rm
odd-odd}}2 I_{\rm odd-odd},
\end{equation}
\begin{equation}
{\Delta E}_{\rm odd} = \frac{{\hbar}^{2}}{2{\cal J}_{\rm odd}}
2I_{\rm odd}.
\end{equation}
Again using Eq. (11) (i.e., the assumption about the contribution of
the decoupled particle to the moment of inertia of the odd-odd
nucleus), the energy intervals in the odd-odd nucleus can be
expressed through the energy intervals in the odd nucleus and the
geometric factor:
\[{\Delta E}_{\rm odd-odd} = \frac{\sin \theta_{\rm odd-odd}}{\sin
\theta_{\rm odd}} {\Delta E}_{\rm odd}=\]
\begin{equation}
=\frac{{\hbar}^{2}}{2{\cal J}_{\rm odd}} \frac{\sin \theta_{\rm
odd-odd}}{\sin \theta_{\rm odd}} 2I_{\rm odd}.
\end{equation}

Thus, the energy intervals of the ``conflicting'' rotational bands
in the odd-odd nuclei are close to those of the  bands in the
neighboring odd nuclei based on the state of a strongly coupled
nucleon entering the ``conflicting'' configuration with the
exception of
the low-lying intervals. The reduction of the low-lying
intervals in the odd-odd nuclei is explained by the different
orientations of the total angular momenta in the odd-odd and odd
nuclei and is a manifestation of the characteristic action of the
decoupled nucleon as a spectator. It is worth noting that a somewhat
different situation is obtained if one compares the energy
intervals of the rotational band in the odd nucleus based on the
state of a decoupled nucleon with the energy intervals of the
ground-state band in the neighboring even-even nucleus. They are
very close also for low-lying states, because the total angular
momentum of the odd nucleus is parallel, in this case, to the
rotational angular momentum of the core (the geometric factor in
(16) is equal to 1).

As an example, the spectra of the rotational bands in the odd-odd
isotopes $^{118,120}$Sb based on the ``conflicting'' state
($\nu$h$_{11/2}\otimes\pi$g$_{9/2})8^{-}$ and those in the odd isotopes
$^{117,119}$Sb \cite{Fos97} based on the state of the
strongly coupled proton g$_{9/2}$ are compared in Fig. 2. For
convenience of comparison, the positions of the band heads in the
odd-odd and odd nuclei have been
shifted in this and other figures. We find that the
the energy intervals are indeed
very close with the exception of the low-lying intervals: those in
the odd-odd isotopes are smaller. The same is observed for the
pairs of isotopes $^{114}$Sb, $^{113}$Sb and $^{116}$Sb,
$^{115}$Sb \cite{Fos97}.

\begin{figure}
\includegraphics[width=7.5cm]{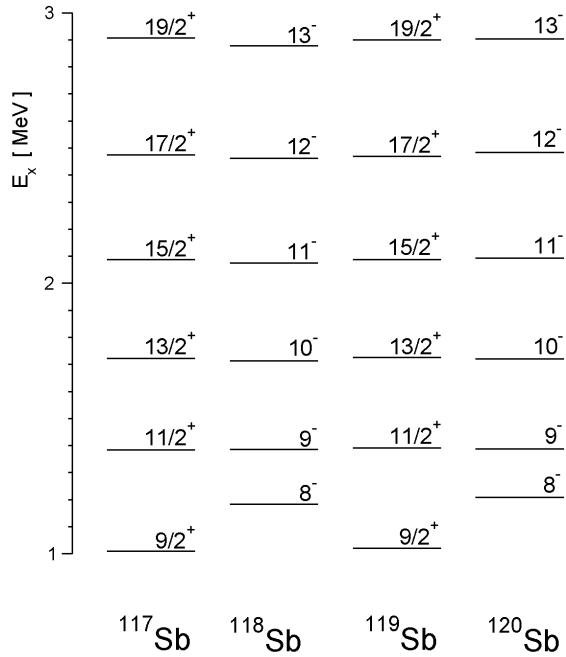}
\caption{ Comparison of the level energies of the rotational
    bands in the pairs of the odd-odd and odd isotopes of Sb. For
        convenience of comparison, the positions of the bands
        are shifted to the same energy
        for 11/2$^+$ and 9$^-$ levels }
\end{figure}

Figure 3 presents the comparison of the rotational band in
$^{130}$La based on the state
$(\nu$g$_{7/2}\otimes\pi$h$_{11/2})8^{-}$ with the strongly
coupled neutron and the decoupled proton and the band in the odd
$^{129}$Ba \cite{Giz77} based on the neutron state
(g$_{7/2})7/2^{+}.$ The same behavior of the energy intervals is
observed also in this case. The spectrum of the band in $^{130}$La
calculated by expression (13) with regard for (16) is also shown
in Fig. 3. The spin dependence of the moment of inertia of the odd
nucleus in expression (16) is taken in the form
\begin{equation}
\frac{\hbar^{2}}{2{\cal J}_{\rm odd}}=\frac{\hbar^{2}}{2{\cal
J}_{0}}+ B[I(I+1)-K^{2}],
\end{equation}
where the parameters ${\cal J}_{0}$ and $B$ were determined from the
analysis of the reference band in $^{129}$Ba. As is seen, the
calculated values fit the experimental data very well including
the characteristic reduction of the low-lying intervals.

\begin{figure}
\includegraphics[width=7.5cm]{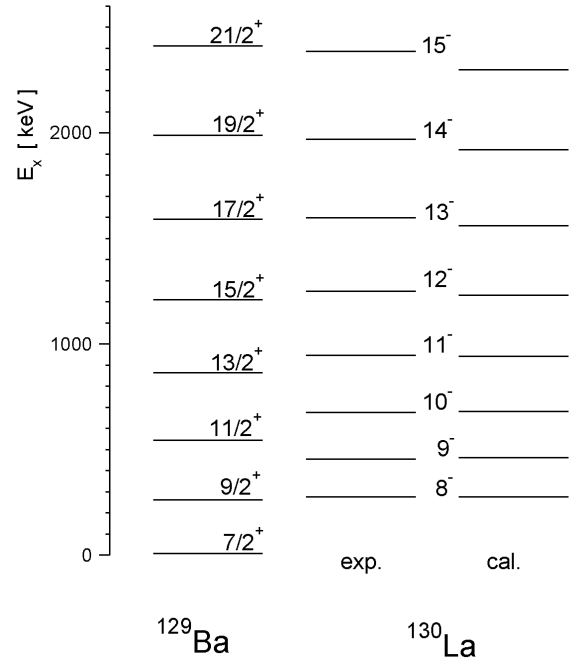}
\vskip-1mm\caption{ Energy spectra of the ``conflicting'' band in
$^{130}$La
    based on the state $(\nu$g$_{7/2}\otimes\pi$h$_{11/2})8^{-}$
    and the band in $^{129}$Ba based on the neutron
       state g$_{7/2}$ strongly coupled to the core.
       The positions of the bands
      are shifted to the same energy
        for 9/2$^+$ and 8$^-$ levels }
\end{figure}

The same picture is observed for the rotational bands based on the
``conflicting'' state $(\nu$h$_{11/2}\otimes\pi$g$_{9/2})8^{-}$ in
$^{120}$Cs and the state $(\pi$g$_{9/2})9/2^{+}$ in $^{119}$Cs
\cite{Gar79}, for which both the experimental and calculated spectra
are shown in Fig. 4.

\begin{figure}
\includegraphics[width=7.0cm]{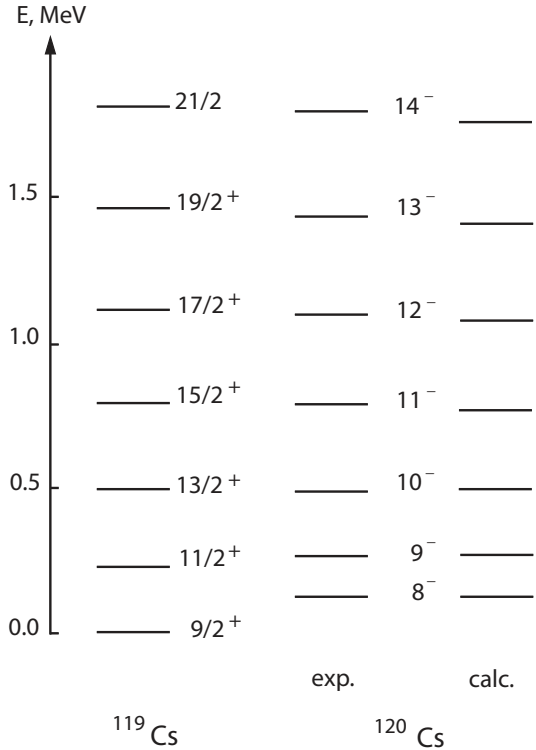}
\vskip-3mm\caption{ Comparison of the energy spectra of the
``conflicting''
    band in $^{120}$Cs based on the state $(\nu$h$_{11/2}
    \otimes\pi$g$_{9/2})8^{-}$ and the band in $^{119}$Cs based
    on the proton state g$_{9/2}$ strongly coupled to the core.
    The positions of the bands
        are shifted to the same energy
        for 13/2$^+$ and 10$^-$ levels }
\end{figure}

Some comments are in place about the choice of $ \langle K \rangle
$ used in the calculations of the level energies in the odd-odd
isotopes of Cs and La. According to the Gallagher--Moszkowski rule
\cite{Gal56}, the odd proton and neutron are coupled to the core
in such a way that the projections of the intrinsic spins on the
symmetry axis are parallel.
 The ``conflicting'' state $(\nu$h$_{11/2}\otimes\pi$g$_{9/2})$ in
the odd-odd isotopes of Cs is formed by the neutron [550]1/2$^{-}$
and the proton [404]9/2$^{+}.$ Therefore, the value $ \langle K \rangle
= 5 $ was used in the calculations. The ``conflicting'' state
$(\nu$g$_{7/2}\otimes\pi$h$_{11/2})$ in the odd-odd isotopes of La
is formed by the neutron [404]7/2$^{+}$ and the proton
[550]1/2$^{-}.$ Therefore, $ \langle K \rangle =3 $ was used.

As nucleons are filling the shell, the odd nucleon initially
aligned along the rotational axis can change its orientation,
because
the proportion between the interactions with the deformation and
the rotation changes (the Fermi level is shifted to the orbitals with
larger values of $\Omega$). A decrease in the alignment of the
nucleon leads simultaneously to an increase of  $\Omega,$ which
results in a change of the geometric factor in Eq.\,(16).
Therefore, the
intervals between the low-lying levels of the ``conflicting''
band are firstly reduced as compared with those of the reference band, and then
the ratios of these intervals get closer to 1 with increase in the
mass number.

As an example, we consider the ratio of the energy intervals of the
rotational bands based on the ``conflicting'' state
$(\nu$h$_{11/2}\otimes\pi$g$_{9/2})8^{-}$ in the odd-odd isotopes
$^{116,118,120}$I and those based on the state of a
strongly coupled proton g$_{9/2}$ in the odd isotopes
$^{117,119,121,123}$I. Such a comparison is given in
Table~\ref{tab:ratio1}.

\begin{table}[b]
\noindent\caption{\label{tab:ratio1}Experimental ratios of the
energy intervals between the levels
   \boldmath$\boldsymbol(\nu$h$_{11/2}\otimes\pi$g$_{9/2})8^{-}$ in the odd-odd isotopes
    of iodine and the band based on the state of a strongly coupled proton
    g$_{9/2}$ in the odd isotopes of iodine }\vskip3mm\tabcolsep9.2pt

\noindent{\footnotesize\begin{tabular}{c c c c }
 \hline \multicolumn{1}{c}
{\rule{0pt}{9pt}$(I_{i}-I_{f})_{\rm odd-odd}/(I_{i}-I_{f})_{\rm
odd}$} & \multicolumn{1}{|c}{$^{116}$I}&
\multicolumn{1}{|c}{$^{118}$I}&
\multicolumn{1}{|c}{$^{120}$I}\\%
\hline%
$(9^{-}-8^{-})/(11/2^{+}-9/2^{+})$&0.732&0.802&0.905 \\%
$(10^{-}-9^{-})/(13/2^{+}-11/2^{+})$&0.887&0.932&1.040 \\%
$(11^{-}-10^{-})/(15/2^{+}-13/2^{+})$&0.938&0.971& 1.051 \\%
$(12^{-}-11^{-})/(17/2^{+}-15/2^{+})$&0.971&0.995&1.060 \\%
\hline
\end{tabular}}
\end{table}

As can be seen, the ratio of energy intervals between the low-lying
levels that are the most sensitive to the orientation of the total
angular momenta
approaches 1 with increase in the mass number in accordance
with our assumption. It approaches also 1, as the
spin increases. The calculated ratios of the energy intervals are given in
Table~\ref{tab:ratio2}. The following values of $ \langle K \rangle
$ were used in the calculations: 5 for $^{116}$I ($ \langle
\Omega_{n} \rangle = 1/2 ,$ $\Omega_{p} = 9/2 $), 6 for $^{118}$I ($
\langle \Omega_{n} \rangle = 3/2 ,$ $\Omega_{p} = 9/2 $), and 7 for
$^{120}$I ($ \langle \Omega_{n} \rangle = 5/2 ,$ $\Omega_{p} = 9/2
$). The variation of the alignment from the maximum value of 5.5 to
the value of 4.9 corresponds to this variation in $ \langle K
\rangle .$

\begin{table}[b]
\noindent\caption{\label{tab:ratio2}Calculated ratios between the
corresponding energy intervals in odd-odd and odd iodine isotopes
}\vskip3mm\tabcolsep9.2pt

\noindent{\footnotesize\begin{tabular}{c c c c }
 \hline \multicolumn{1}{c}
{\rule{0pt}{9pt}$(I_{i}-I_{f})_{\rm odd-odd}/(I_{i}-I_{f})_{\rm
odd}$} & \multicolumn{1}{|c}{$^{116}$I}&
\multicolumn{1}{|c}{$^{118}$I}&
\multicolumn{1}{|c}{$^{120}$I}\\%
\hline%
$(9^{-}-8^{-})/(11/2^{+}-9/2^{+})$&0.691&0.772&0.914 \\%
$(10^{-}-9^{-})/(13/2^{+}-11/2^{+})$&0.833&0.902&1.010 \\%
$(11^{-}-10^{-})/(15/2^{+}-13/2^{+})$&0.898&0.954&1.040 \\%
$(12^{-}-11^{-})/(17/2^{+}-15/2^{+})$&0.933&0.979&1.044 \\%
\hline
\end{tabular}}
\end{table}

The small deviations from experiment of the calculated first four
intervals in $^{116,118,120}$I, the accuracy of the calculations
for the energy spectra in the odd-odd nuclei $^{130}$La,
$^{120}$Cs, $^{170}$Ta, and the analysis of the energy spectra of
the isotones with N=75 show that the phenomenological
approximation used here describes the rotational bands in the
odd-odd nuclei based on the ``conflicting'' states in a
satisfactory way. The actual quantum state of an odd-odd nucleus
on a microscopic level can be a rather complicated superposition
of simple states, and the substitution of the quantum-mechanical
angular-momentum operator by its classical expectation value does
not take this circumstance into account. But the observed
similarity of the bands in the odd and odd-odd nuclei indicates
that this superposition is not changed much in going from the odd
nucleus to the `conflicting` state in the odd-odd nucleus, thus
justifying the simple interpretation.

Let us return to the effects of the residual interaction. The
configurations of the odd proton and neutron must change under
rotation. The nucleon firstly oriented along the symmetry axis
changes its orientation under the influence of the Coriolis
interaction \cite{Kre77}, and the term $B$ in eq. (3) can differ
from zero for high-spin states. Moreover, the nondiagonal matrix
elements between different configurations can become important
\cite{Tok77}. To accurately calculate these effects, one needs to
know the wave functions of the model used, by sacrificing
the clearity and the simplicity. But the good agreement with experiment of
the simple calculations presented for the whole band and the
similarity of the bands in the odd-odd and in the corresponding
odd nuclei (apart from the lowest levels) indicates that the residual
interaction of the odd nucleons has only a small
 effect on the structure  of the rotational band in the case of the
 ``conflicting''
coupling. All effects of the Coriolis interaction are the same in
the ``conflicting'' band and in the reference band of the
neighboring odd nucleus. The problem of the residual interaction
has been previously discussed in connection with the structure of
the rotational bands in odd-odd nuclei in the Tl-region
\cite{Yad78,Tre80}. It was found that the main features of these
bands can be explained by the Coriolis interaction without taking
into account effects of the residual interaction \cite{Kre80,Tre80}.
Moreover if taken into account the residual interaction gives a
signature-dependent contribution which has the opposite phase as
needed to explain the staggering. This also supports our confidence
that the residual interaction does not influence the structure of
the collective bands in the case of `conflicting` coupling.

\section{Structure of the Collective Bands in Odd-Odd
Nuclei for ``Weakly Conflicting'' Coupling}

\begin{figure}
\includegraphics[width=7.5cm]{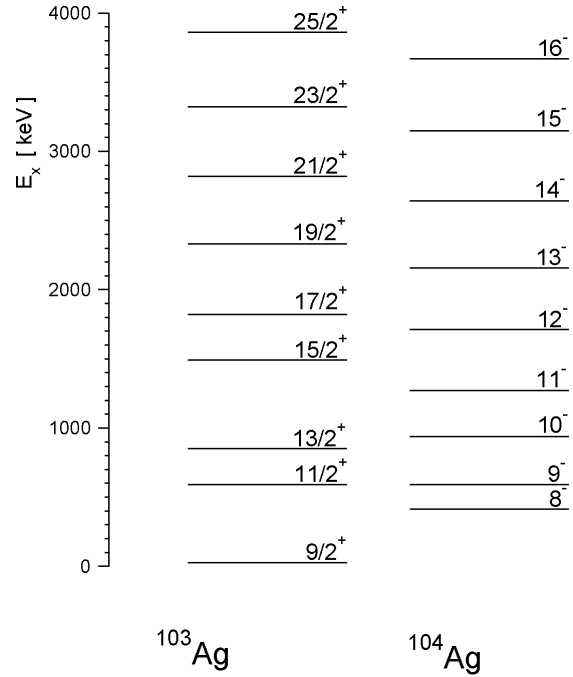}
\vskip-3mm\caption{ Comparison of the energy spectra of the
rotational
    bands in the odd-odd nucleus $^{104}$Ag and in the odd
    nucleus $^{103}$Ag. The positions of the bands
         are shifted to the same energy
        for 11/2$^+$ and 9$^-$ levels }\label{fig:Ag}
\end{figure}

In the  ``conflicting'' coupling considered, the Fermi level for
a strongly coupled nucleon is close to the orbital with the
largest value of the angular momentum projection on the symmetry
axis, $\Omega=j.$ The Coriolis interaction is weak in this case
and can be neglected. If, however, the Fermi level for a
strongly coupled nucleon is situated among the intermediate values
of $\Omega,$ then the Coriolis interaction can be large enough to
perturb the rotational spectrum. This situation may be termed as
``weakly conflicting'' coupling. For odd nuclei, particularly
large perturbations are observed in the case of an admixture of
orbitals with $\Omega=1/2.$ The sign-changing diagonal matrix
elements of the Coriolis interaction must be taken into account.
This leads to the following shift of the level with spin $I$
\cite{Jai89}:
\begin{equation}
\Delta E=-A(-)^{I+1/2}(I+1/2)a.
\end{equation}
Here, $a$ is the decoupling parameter determined by the
coefficients of the expansion of the intrinsic wave function in
terms of the eigenfunctions of the operator $\hat{j}.$ This
results in a characteristic doublet character of the rotational
bands. The larger the decoupling parameter, the stronger the
perturbation of the rotational bands. Figure \ref{fig:Ag} shows
the rotational band in $^{103}$Ag \cite{Gos95} based on the proton
state g$_{9/2}.$ Let us limit ourselves to the single-orbit
approximation \cite{Pav81}. As the decoupling parameter is
proportional to $ (-)^{j+1/2},$ this leads to a negative value of
$a(\pi$g$_{9/2}).$ Therefore, the levels with the spins
$I_{0}\!+\!1, I_{0}\!+\!3,$~\ldots ($I_{0}$ is the spin of the
band head) are shifted up and the levels with the spins
$I_{0}\!+\!2, I_{0}\!+\!4,$~\ldots are shifted down.

For odd-odd nuclei, the sign-changing  diagonal matrix elements
equivalent to (18) do not exist. Nevertheless, the examination of the
rotational part of Hamiltonian (2) shows that a similar role
is played by the non-diagonal matrix elements of the Coriolis
interaction coupling the states with $K=1$ and $K=0$
\cite{Jai89,Gos95}:
\[
\langle K=1\mid \hat{H}_{c} \mid K=0 \rangle =
\]
\begin{equation}
=
-A[a_{p}+(-)^{I+1}a_{n}][I(I+1)]^{1/2}.
\end{equation}

Let us consider the ``weakly conflicting'' state
$(\nu$h$_{11/2}\otimes\pi$g$_{9/2})$ in the odd-odd isotopes of Ag,
where the proton Fermi-level lies between the orbitals
[413]7/2$^{+}$ and [404]9/2$^{+}$ \cite{Tre80}. As
$a(\pi$g$_{9/2})<0,$ $a(\nu$h$_{11/2})>0,$
and
 $a_{n}>a_{p}$ (the neutron is decoupled),
the sign of the matrix elements (19) is determined by the term  \(
(-)^{I+1}a_{n}[I(I+1)]^{1/2} \). Therefore, the situation is
opposite to that in the odd isotopes of Ag: the levels with spins
$I_{0}\!+\!1, I_{0}\!+\!3,$ \ldots are shifted down, while the
levels with
 spins $I_{0}\!+\!2, I_{0}\!+\!4,$ \ldots are shifted up.

The reduction of the energy intervals between the low-lying
levels in $^{104}$Ag as compared with the corresponding levels in the
odd $^{103}$Ag can be explained in a similar way as for the case
of the extreme ``conflicting'' coupling, namely by the different
 orientations of the total angular momenta.

\section{Conclusion}

The approximation based on the quasiclassical limit of the model
``axial rotor + two quasiparticles'' allowed us to establish that
the rotational band in odd-odd nuclei based on the ``conflicting''
state can be considered as the rotational excitations of the
neighboring odd nucleus, in which the odd strongly coupled particle
is the one entering the ``conflicting'' configuration in the odd-odd
nucleus. However, the decoupled particle in the odd-odd nucleus
cannot be considered simply as a spectator, as in the decoupled band
in the odd nucleus. Its presence leads to a change of the
orientation of the total angular momentum in the odd-odd nucleus as
compared with that in the odd nucleus and to a modification of the
moment of inertia, which can be expressed by a geometric factor.
This is the reason for the reduction of the low-lying
energy-intervals between the band levels in an odd-odd nucleus
relative to those in an odd nucleus. For the high-lying states, the
bands are very similar, because the geometric factor
approaches~1.\looseness=1

\vskip3mm The authors would like to thank Profs. V.M.~Ko\-lo\-miets
from Institute for Nuclear Research, Kyiv, J.~Kva\-sil from Charles
University, Prague, and H.~Wol\-ter from University Munich for the
critical reading of the manuscript and useful suggestions.

\rezume{%
КОНФЛІКТНИЙ ЗВ'ЯЗОК НЕСПАРЕНИХ НУКЛОНІВ\\ І СТРУКТУРА КОЛЕКТИВНИХ
СМУГ\\ В НЕПАРНО-НЕПАРНИХ ЯДРАХ}{О.І. Левон, О.А. Пастернак}
{Розглянуто конфліктний зв'язок  неспарених нуклонів в
 непарно-непарних ядрах. Запропоновано просте пояснення
 стискання нижніх енергетичних рівнів ротаційних смуг в
 непарно-непарних ядрах у випадку конфліктного зв'язку непарних
 протона і нейтрона порівняно з ротаційними смугами,
 побудованими на стані сильно зв'язаного нуклона в сусідньому
 непарному ядрі, тим же, що входить в конфліктний стан.}

\end{document}